\documentclass[12pt, preprint]{aastex}
\begin{document}
\title{An Analytic Formula for the Supercluster Mass Function}
\author{Seunghwan Lim\altaffilmark{1}, Jounghun Lee\altaffilmark{2}}
\altaffiltext{1}{Department of Astronomy,University of Massachusetts, LGRT-B 619E, 710 North Pleasant Street, 
Amherst, MA 01003-9305, USA; slim@astro.umass.edu}
\altaffiltext{2}{Astronomy Program, Department of Physics and Astronomy, FPRD, 
Seoul National University, Seoul 151-747, Korea; jounghun@astro.snu.ac.kr}
\begin{abstract}
We present an analytic formula for the supercluster mass function which is constructed by modifying the extended 
Zel'dovich model for the halo mass function. The formula has two characteristic parameters whose best-fit values 
are determined by fitting to the numerical results from N-body simulations for the standard $\Lambda$CDM 
cosmology. The parameters are found to be independent of redshifts and robust against variation of the key 
cosmological parameters. Under the assumption that the same formula for the supercluster mass function is 
valid for non-standard cosmological models, we show that the relative abundance of the rich superclusters 
should be a powerful indicator of any deviation of the real universe from the prediction of the standard $\Lambda$CDM 
model.  
\end{abstract}
\section{INTRODUCTION}
\label{sec:intro}

The gravitational aggregates of (a few to hundreds of) galaxy clusters are 
called {\it the superclusters} which are marginally bound systems.
Although quite rare in the local universe, the superclusters are believed to be 
common phenomena on the scales larger than $100$ Mpc. The nearby Virgo 
cluster as well as the Local Group where our Milky Way resides also belongs to 
the Local Supercluster that contains more than $100$ member clusters 
\citep[][and references there in]{tully82}.
   
In the standard $\Lambda$CDM ($\Lambda$+cold dark matter) universe, 
the formation of the superclusters at the present epoch represents the grand finale of the 
hierarchical merging events. No bound objects could form on mass scale larger 
than that of the present rich superclusters in the future due to the 
anti-gravitational effect of the cosmological constant ($\Lambda$). That 
is, the rich superclusters observed at the present epoch will end up as isolated 
massive clusters in the future when $\Lambda$ becomes progressively more 
dominant \citep[e.g., see][]{NL03,KE05,busha-etal05,araya-etal09}.

If the dark energy were not $\Lambda$ or if the large-scale gravity deviated 
from the general relativity (GR), the superclusters could meet a different 
fate. For example, in QCDM (Quintessence+CDM) models, more massive objects 
than the rich superclusters would form through the large scale clustering of the 
Quintessence scalar field \citep[see][and references therein]{caldwell-etal98}. 
In some modified gravity scenarios where the universe has no anti-gravitational 
dark energy \citep[for a review, see][]{mg_review12}, nothing would prevent the superclusters 
from assembling into larger scale objects.  Henceforth, the abundance of rich superclusters 
might be a powerful indicator of any deviation of the real universe from the prediction of the 
standard $\Lambda$CDM model.

There are two advantages that the number count of rich superclusters 
has as a cosmological probe over that of the clusters. First of all, the 
rich superclusters are larger and rarer on average than the clusters, and thus 
their abundance should be more sensitive to the background cosmology.  
The other advantage is that since the superclusters are still in the quasi-linear 
regime, their formation process would be less affected by the complicated nonlinear 
effect and thus their mass function (defined as the number density of the superclusters 
per unit volume as a function of mass) may be easier to model theoretically. 

It was \citet{oguri-etal04} who for the first time attempted to find an analytic 
expression for the supercluster mass function in the framework of the standard 
Press-Schechter theory \citep{PS74}. Pioneering as it was, 
their work was based on a few unjustified assumptions, which resulted in the failure of 
their model in matching the N-body results (M. Oguri in private communication). 
\citet{YF11} determined numerically the mass function of the supercluster-like filaments 
with the help of large N-body simulations and showed that the standard excursion set theory 
is incapable of reproducing the numerical results. Although they claimed that the incorporation 
of the "peak-exclusion effect" could lead the excursion set mass function to  
match qualitatively the numerical results,  their model could not pull it off quantitatively.
 
Our goal here is to find an efficient analytic formula for the supercluster mass function by modifying the 
extended Zel'dovich (EZL) model for the halo mass function which was constructed in our previous work 
\citep{LL13}. The EZL model is based on the formalism originally developed by \citet{jedamzik95} and characterized 
by three parameters expressed in terms of the thresholds of the linear shear eigenvalues. 
As mentioned clearly in \citet{LL13}, the EZL model is not a physical model but a fitting formula 
whose characteristic parameters have to be determined empirically by adjusting the model to numerical 
results. Nevertheless, it turned out that the best-fit parameters of the EZL model are independent of redshift,  
having the same values even when the background cosmology changes. Furthermore, the EZL formula 
for the halo mass function was shown to agree with the N-body results better than the other analytic 
formula suggested so far \citep[e.g.,][]{ST99,tinker-etal08,PPH10}, which encourages us to use it as a basic 
framework for the analytic construction of the supercluster mass function.
  
This Paper is  composed of five sections whose contents are outlined as  
follows: In section \ref{sec:model} we provide a brief review of the EZL model and 
describe how the EZL model is modified to construct a new formula for the supercluster 
mass function.  In section \ref{sec:test} we compare our model with the numerical  
results from three different N-body simulations. In section \ref{sec:rich} we explore the possibility 
of using the relative abundance of rich superclusters as a new cosmological probe.
In section \ref{sec:con} we discuss the results and assess its cosmological importance.

\section{CONSTRUCTING A SUPERCLUSTER MASS FUNCTION}\label{sec:model}

\subsection{Summary of the EZL model}

The EZL model for the differential mass function of bound halos, $dN/dM$, is expressed by the following integral 
equation
\begin{equation}
\label{eqn:nft}
F(\lambda_{1c},\lambda_{2c},\lambda_{3c};M)
=\int_{M}^{\infty}dM^{\prime}{M^{\prime}\over{\bar\rho}}\frac{dN}{dM^{\prime}}
P(M,M^{\prime})\ ,
\end{equation}
which look similar to the formula originally developed by \citet{jedamzik95}. The key difference of 
equation (\ref{eqn:nft}) from the original Jedamzik formula is that the characteristic parameters of the EZL 
model are expressed in terms of the thresholds (i.e., lower limits) of three eigenvalues of the linear deformation 
tensor, $\lambda_{1c},\ \lambda_{2c},\ \lambda_{3c}$.

The left-hand side in equation (\ref{eqn:nft}) is equal to the cumulative probability that the linear shear 
eigenvalues $\lambda_{1},\ \lambda_{2},\ \lambda_{3}$ (with $\lambda_{1}\ge\lambda_{2}\ge\lambda_{3}$) 
on the mass scale $M$ exceed their thresholds: 
\begin{eqnarray}
\label{eqn:flambda1}
F(M) &=& 
P[\lambda_{1}\ge\lambda_{1c},\lambda_{2}\ge\lambda_{2c},\lambda_{3}\ge\lambda_{3c}|\sigma(M)]\, , \\
\label{eqn:flambda2}
&=& \int^{\infty}_{\lambda_{1c}}d\lambda_{1}\, 
\int^{\lambda_{1}}_{\lambda_{2c}}d\lambda_{2}\, 
\int^{\lambda_{2}}_{\lambda_{3c}}d\lambda_{3}\, 
p[\lambda_{i};\sigma(M)], 
\end{eqnarray}
where $\sigma(M)$ is the rms  fluctuation of the linear density contrast  smoothed by a top-hat filter on the mass scale 
of $M$. Equation (\ref{eqn:flambda2}) can be straightforwardly calculated from the joint probability density distribution of 
the linear shear eigenvalues derived by \citet{dor70}
\begin{equation}
\label{eqn:lam}
p(\lambda_{1},\lambda_{2},\lambda_{3}) 
= \frac{3375}{8\sqrt{5}\pi\sigma^{6}}
\exp\left[-\frac{3}{\sigma^{2}}\left(\sum_{i=1}^{3}\lambda_{i}\right)^{2} + 
 \frac{15}{2\sigma^{2}}\sum_{i>j}\lambda_{i}\lambda_{j}\right]
\vert\Pi_{i >j}(\lambda_{i}-\lambda_{j})\vert,
\end{equation}

The core quantity in the right-hand side of equation (\ref{eqn:nft}) is the conditional probability $P(M,M^{\prime})$ 
defined as 
\begin{equation}
\label{eqn:pmm}
P(M,M^{\prime})\equiv \int_{\lambda_{1c}}^{\infty}d\lambda_{1}
\int_{\lambda_{2c}}^{\lambda_{1}}d\lambda_{2}
\int_{\lambda_{3c}}^{\lambda_{2}}d\lambda_{3}\,
p_{c}(\lambda_{i} | \lambda^{\prime}_{i}=\lambda_{ic})\,
\end{equation}
where $p_{c}(\lambda_{i} | \lambda^{\prime}_{i}=\lambda_{ic})$ is the conditional joint probability density evaluated as  
\begin{equation}
\label{eqn:cpll}
p_{c}(\lambda_{1},\lambda_{2},\lambda_{3} | \lambda^{\prime}_{1}=\lambda_{1c},
\lambda^{\prime}_{2}=\lambda_{2c}, \lambda^{\prime}_{3}=\lambda_{3c})=
\frac{p(\lambda_{1},\lambda_{2},\lambda_{3}, \lambda^{\prime}_{1}=\lambda_{1c},
\lambda^{\prime}_{2}=\lambda_{2c}, \lambda^{\prime}_{3}=\lambda_{3c})}{p(\lambda^{\prime}_{1}=\lambda_{1c},
\lambda^{\prime}_{2}=\lambda_{2c}, \lambda^{\prime}_{3}=\lambda_{3c})}\, .\
\end{equation}
where $\{\lambda^{\prime}_{i}\}_{i=1}^{3}$represent the shear eigenvalues on some larger mass scale 
$M^{\prime}\ge M$.

Equations (\ref{eqn:pmm})-(\ref{eqn:cpll}) require to have  the joint probability density distribution of 
the shear eigenvalues on two different mass scales, $M$ and $M^{\prime}$, which have been already 
analytically found in the ingenious works of \citet{desjacques08} and \citet{DS08}: 
\begin{eqnarray}
\label{eqn:pjoint}
p(\lambda_{1},\lambda_{2},\lambda_{3}, \lambda^{\prime}_{1},
\lambda^{\prime}_{2}, \lambda^{\prime}_{3})&=&\frac{15^6}{320\pi^2\sigma^6\sigma^{\prime6}}
(1-\gamma^2)^{-3}w(\beta\epsilon_-,\epsilon_{\lambda^{\prime}}, 
\epsilon_{\lambda})\mathrm{e}^{-Q+\beta\epsilon_+}\times\, \\
&&\vert\Pi_{i> j}(\lambda_{i}-\lambda_{j})
\Pi_{i> j}(\lambda^{\prime}_{i}-\lambda^{\prime}_{j})\vert.
\end{eqnarray}
\setlength\arraycolsep{0pt}
\begin{eqnarray}
\gamma &=&\frac{1}{2\pi^2\sigma\sigma^\prime}
\int_0^\infty{d(\ln k)k^3P(k)W(k;M)W(k;M^\prime)}\, , \nonumber \\
\beta &=&\frac{15\gamma}{2(1-\gamma^2)}\, , \nonumber \\
Q &=&\frac{3}{4(1-\gamma^2)}\left\{5[\mathrm{tr}(\lambda^{\prime2})+\mathrm{tr}
(\lambda^{\prime2})]-(\mathrm{tr}\lambda^\prime)^2-(\mathrm{tr}\lambda)^2
+2\gamma(\mathrm{tr}\lambda^\prime)(\mathrm{tr}\lambda)\right\}\, , \nonumber \\
w &=&
\frac{\mathrm{e}^{-\beta\epsilon_-}}{2\pi}\int_0^1{dr}\int_0^{2\pi}
{d\varphi\exp\Bigg[\frac{3\beta\epsilon_-}{4}g\Bigg] 
I_0\Bigg[\frac{3\beta\epsilon_-\epsilon_\lambda}{4}\sqrt{h}\Bigg]}\, ,
\end{eqnarray}
where $g$ and $h$ are given as
\begin{eqnarray}
g &=&1+r^2+\epsilon_{\lambda^\prime}(1-r^2)\cos(2\varphi)\, , \\
h &=& g^2-4(1-\epsilon_{\lambda^\prime}^2)r^2\, , 
\end{eqnarray} 
and $\sigma^\prime\equiv\sigma(M^\prime)$, $\sigma\equiv\sigma(M)$, and 
\begin{eqnarray}
\label{eqn:tr}
&&\mathrm{tr}\lambda^\prime = \sum_i{(\lambda^\prime/\sigma^\prime)},\  
\mathrm{tr}\lambda=\sum_i{(\lambda/\sigma)}, \\
\label{eqn:trs}
&&\mathrm{tr}(\lambda^{\prime2}) = \sum_i{(\lambda^\prime/\sigma^\prime)^2},\ 
\mathrm{tr}(\lambda^2)=\sum_i{(\lambda/\sigma)^2},\ \\
\label{eqn:ep+}
&&\epsilon_+=\frac{1}{3}(\mathrm{tr}\lambda^\prime)(\mathrm{tr}\lambda),\ \\
\label{eqn:ep-}
&&\epsilon_-=\frac{1}{3}
(\mathrm{tr}\lambda^\prime-3\lambda_3^\prime/\sigma^\prime)
(\mathrm{tr}\lambda-3\lambda_3/\sigma),\ \\
\label{eqn:epl}
&&\epsilon_{\lambda^\prime}=(\lambda_1^\prime-\lambda_2^\prime)/
(\mathrm{tr}\lambda^\prime-3\lambda_3^\prime/\sigma^\prime),\ \\
\label{eqn:eplp}
&&\epsilon_\lambda=(\lambda_1-\lambda_2)/(\mathrm{tr}\lambda-3\lambda_3/\sigma).
\end{eqnarray}

Rewriting equation (\ref{eqn:nft}) as a discrete matrix equation and solving it through equations 
(\ref{eqn:flambda1})-(\ref{eqn:eplp}), \citet{LL13} obtained the halo mass function, $dN/dM$ and determined 
the best-fit values of $\{\lambda_{1c},\ \lambda_{2c},\ \lambda_{3c}\}$ by adjusting the EZL model to the numerical 
results.  As mentioned in section \ref{sec:intro}, the EZL mass function of bound halos turned out to be in excellent 
agreements with N-body results and the best-fit values of  $\{\lambda_{1c},\ \lambda_{2c},\ \lambda_{3c}\}$ 
were shown to be independent of redshifts and the background cosmology. 

\subsection{Modification of the EZL model}

To construct a new formula for the supercluster mass function by modifying the EZL model, we first modify 
the LHS of equation (\ref{eqn:nft}) into 
\begin{eqnarray}
\label{eqn:new_flambda1}
F(M) &\propto& P[\lambda_{1}\le\lambda_{1c},\lambda_{2}\ge\lambda_{2c},\lambda_{3}\ge 0|\sigma(M)]\, , \\
\label{eqn:new_flambda2}
&\propto& \int^{\lambda_{1c}}_{\lambda_{2c}}d\lambda_{1}\, 
\int^{\lambda_{1}}_{\lambda_{2c}}d\lambda_{2}\, 
\int^{\lambda_{2}}_{0}d\lambda_{3}\, 
p[\{\lambda_{i}\};\sigma(M)], 
\end{eqnarray}
where the largest shear eigenvalue, $\lambda_{1}$, has an upper limit rather than a lower limit unlike in 
equation (\ref{eqn:flambda2}) and the lower limit of the smallest eigenvalue, $\lambda_{3}$, is fixed at zero.  
Our new formula for the supercluster mass function has two characteristic parameters, $\lambda_{1c}$ and 
$\lambda_{2c}$, which represent the lower and the upper limit sof the largest and the second to the largest 
eigenvalues of the linear deformation tensor, respectively. 

\begin{figure}
\includegraphics[scale=0.9]{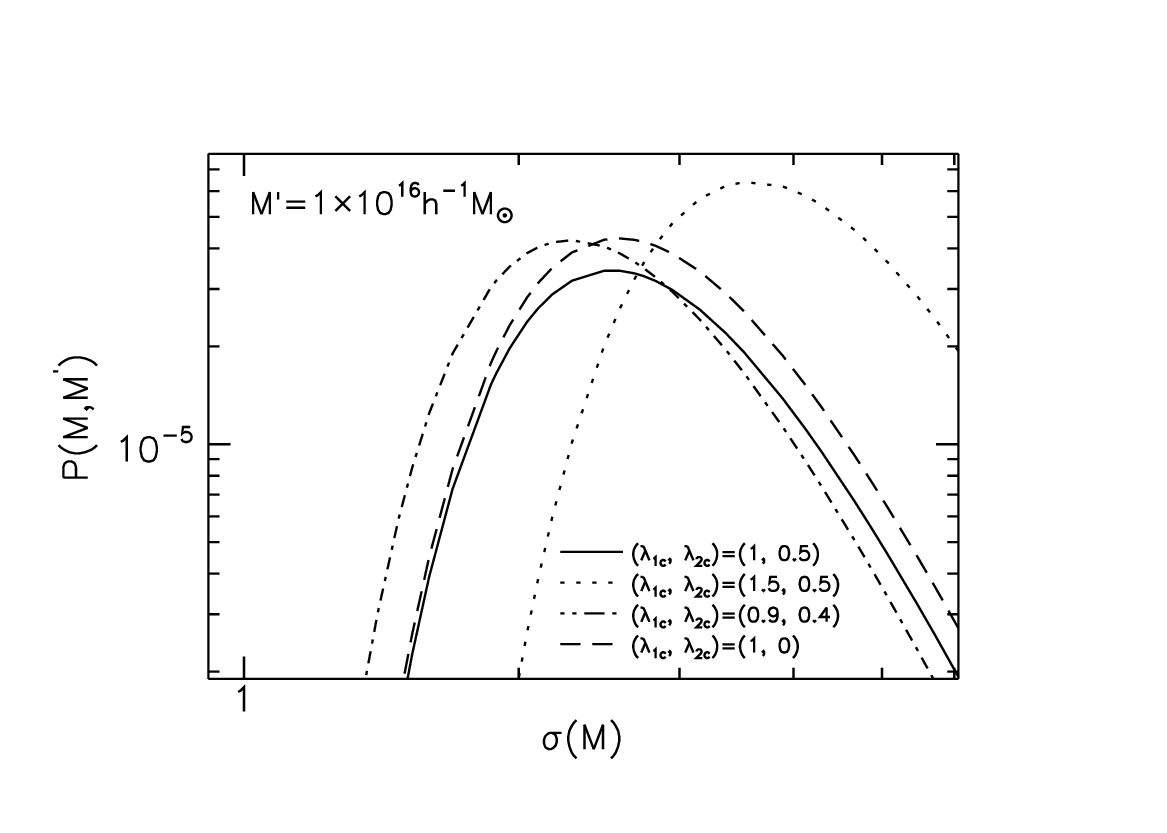}
\caption{Conditional probability, $P(M,M^{\prime})$ in equation (\ref{eqn:new_pmm}), 
for four different cases of the characteristic parameters.}
\label{fig:pmm}
\end{figure}

The conditional probability, $P(M,M^{\prime})$, in the RHS of equation (\ref{eqn:nft})  is accordingly modified into 
\begin{equation}
\label{eqn:new_pmm}
P(M,M^{\prime})=\int_{\lambda_{2c}}^{\lambda_{1c}}d\lambda_{1}
\int_{\lambda_{2c}}^{\lambda_{1}}d\lambda_{2}
\int_{0}^{\lambda_{2}}d\lambda_{3}\,
p_{c}(\lambda_{i} | \lambda^{\prime}_{i}=\lambda_{ic})\,
\end{equation}
Figure \ref{fig:pmm} plots this conditional probability distribution versus the rms density fluctuation for four different 
cases of the characteristic parameters, $\lambda_{1c}$ and $\lambda_{2c}$. As can be seen, the conditional 
probability distribution, $P(M,M^{\prime})$, sensitively changes as the values of the characteristic parameters 
change. 

As done in \citet{LL13}, rewriting equation (\ref{eqn:nft}) as a discrete matrix equation and solving it for 
$dN/dM$ through equations (\ref{eqn:new_flambda1})-(\ref{eqn:new_pmm}), one can obtain the differential mass 
function of the marginally bound superclusters, $dN/dM$.
It is, however, worth noting here that unlike the case of bound halos where the EZL formula yields an 
automatically normalized mass function, for the case of the superclusters the modified EZL formula has to be
renormalized to satisfy the following condition. 
\begin{equation}
\label{eqn:norm}
\int_{M_{\rm c,th}}dM\ \frac{dN}{dM} = \frac{N_{\rm T}}{V_{\rm T}}\, ,
\end{equation}
where $N_{\rm T}$ denotes the total number of superclusters  with mass larger than $M_{\rm c,th}$ 
found in the volume of $V_{\rm T}$ (see section \ref{sec:test}). Equation (\ref{eqn:norm}) will be used to 
determine  the value of the proportionality constant (i.e., normalization factor) in equation (\ref{eqn:new_flambda1}). 
This renormalization step is necessary since not all initial regions  would form marginally bound 
superclusters in the end.  

We would like to state explicitly here that this modified EZL formula for the supercluster 
mass function is not a physical model and the characteristic parameters, 
$\{\lambda_{1c},\ \lambda_{2c},\ \lambda_{3c}\}$, are not related to any underlying dynamics but should be
determined by numerical experiments. In other words, we note only empirically that the modification of 
Equation (\ref{eqn:flambda2}) into Equation (\ref{eqn:new_flambda2}) works for the case of the 
supercluster mass function.

\section{NUMERICAL TESTS}
\label{sec:test}

To test our model for the supercluster mass function against the numerical results, we utilize 
three different N-body simulations: the Millennium \citep{millennium}, the CoDECS \citep{codecs} 
and the MICE \citep{mice} simulations, all of which ran for a flat $\Lambda$CDM 
cosmology but with slightly different cosmological parameters 
\footnote{As for the CoDECS, the simulations ran not only for a flat $\Lambda$CDM 
cosmology but also for various coupled dark energy models \citep{codecs}. 
Here the simulation data only for a $\Lambda$CDM model is used.}. 
Table \ref{tab:simulation} lists the linear size of the simulation box, mass 
resolution,  total number of particles, values of the three key cosmological 
parameters, and the halo-finding algorithm used for the three simulations. 
The Friends-of-Friends (FoF) algorithm with linking length of 
$0.2\bar{l}_{p}$ (where $\bar{l}_{\rm p}$ is the mean particle separation) 
was used in all of the simulations to find the bound groups of dark 
matter particles. See the above three literatures for the full descriptions of 
the simulations and the halo-identification procedure.

The publicly available halo catalogs from each simulation provide such 
information on the resolved halos as their mass, positions, velocities and so 
forth at various redshifts. Analyzing the catalogs from each simulation at 
three different redshifts ($z=0,\ 0.5,\ 1$ for the cases of the Millennium and 
MICE catalogs while $z=0,\ 0.44,\ 1$ for the case of the CoDECS catalog), 
we first construct a mass-limited sample of the clusters from each  
catalog which includes only those halos with masses larger than a threshold value,  
$M_{\rm c, th}=10^{13}\,h^{-1}M_{\odot}$. For the case of the MICE halo catalogs, 
however, a slightly higher threshold value, $M_{\rm c, th}=3.4\times 10^{13}\,h^{-1}M_{\odot}$, 
is used since all of the halos in the MICE catalogs have masses larger than this higher 
threshold value.

Then, we identify the superclusters as the clusters of clusters by applying the FoF 
algorithm to the clusters in each mass-limited sample at each redshift. 
Following the conventional criterion suggested in the previous literatures 
\citep[e.g.,][]{KE05,wray-etal06,LE07}, we set the linking length of the FoF algorithm for the 
supercluster identification at $\bar{l}_{c}/3$ where $\bar{l}_{c}$ is the mean cluster 
separation.The mass of each identified supercluster, $M$, is measured as the sum of the 
masses of its member clusters. The total number and mean mass of the 
superclusters from the three simulations at three different redshifts are 
listed in Tables \ref{tab:sample_z0}-\ref{tab:sample_z1}. Note that among the identified 
superclusters are included those which have only one member cluster (i.e., isolated clusters).

Binning the supercluster mass in the logarithmic scale, $\ln M$, and counting the number of the 
superclusters belonging to each differential mass bin, $[\ln M,\ \ln M + d\ln M]$, we determine the 
number density of the superclusters per unit volume, $dN/d\ln M$, as a function of $M$ 
at each redshift.  To estimate the errors associated with the measurement of 
$dN/d\ln M$, we also perform the Jack-knife analysis: Dividing the superclusters 
into eight Jackknife subsamples, we determine $dN/d\ln M$ for each Jackknife 
subsample and calculate the one standard deviation scatter among the eight 
Jack-knife subsamples as the errors associated with the measurement of 
$dN/d\ln M$. 

Now that the numerical results are all obtained, we want to compare them with the analytic formula 
constructed in section \ref{sec:model}. 
With the help of $\chi^{2}$ minimization scheme, we first fit the numerical results from the Millennium simulation 
to the analytic supercluster mass function and determine the best-fit values of the two parameters to be 
$\lambda_{1c}=0.5$ and $\lambda_{2c}=0.5$, respectively, at $z=0$. Then, we examine whether or not the 
analytic model with the same best-fit values of the parameters still work at higher redshifts. For this comparison, 
the cosmological parameters are set at the same values that were used for the Millennium 
simulation when the analytic formula is calculated and the power spectrum of the standard $\Lambda$CDM  
is evaluated with the help of the CAMB code \citep{camb}.

\begin{figure}
\includegraphics[scale=0.85]{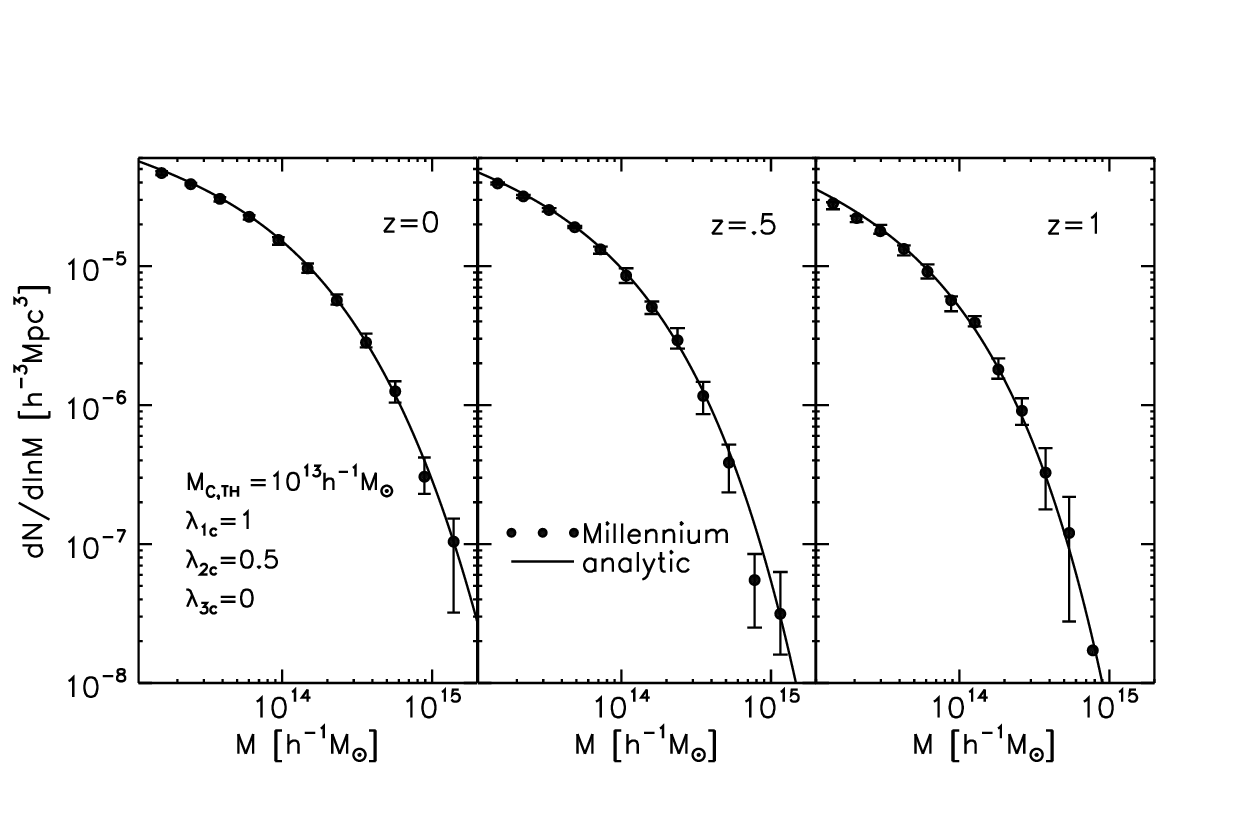}
\caption{Comparison of the supercluster mass functions (solid line) 
with the numerical results from the Millennium simulations (dots) 
at three different redshifts. In each panel the errors represent the one standard 
deviation scatter among eight Jackknife resamples. The mass threshold for the 
supercluster membership, $M_{\rm c,th}$ is set at $10^{13}\,h^{-1}M_{\odot}$.}
\label{fig:mill}
\end{figure}

Figure \ref{fig:mill} plots the numerical results of the supercluster mass function from the Millennium 
simulations (solid dots) and compares them with the analytic models (solid line) with 
the best-fit parameters at $z=0,\ 0.5$ and $1$ in the left, middle and right panel, 
respectively.  The supercluster mass function at higher redshifts can be readily evaluated just by 
substituting $\sigma(M,z)\equiv D(z)\sigma(M)$  for $\sigma(M)$ where $D(z)$ is the 
linear growth factor normalized to be unity at $z=0$. The functional form of $D(z)$ for 
a flat $\Lambda$CDM cosmology is given in \citet{lahav-etal91}.
As can be seen in Figure \ref{fig:mill}, the analytic supercluster mass function with the same best-fit 
parameters agree excellently with the numerical results at all three redshifts. 

\begin{figure}
\includegraphics[scale=0.85]{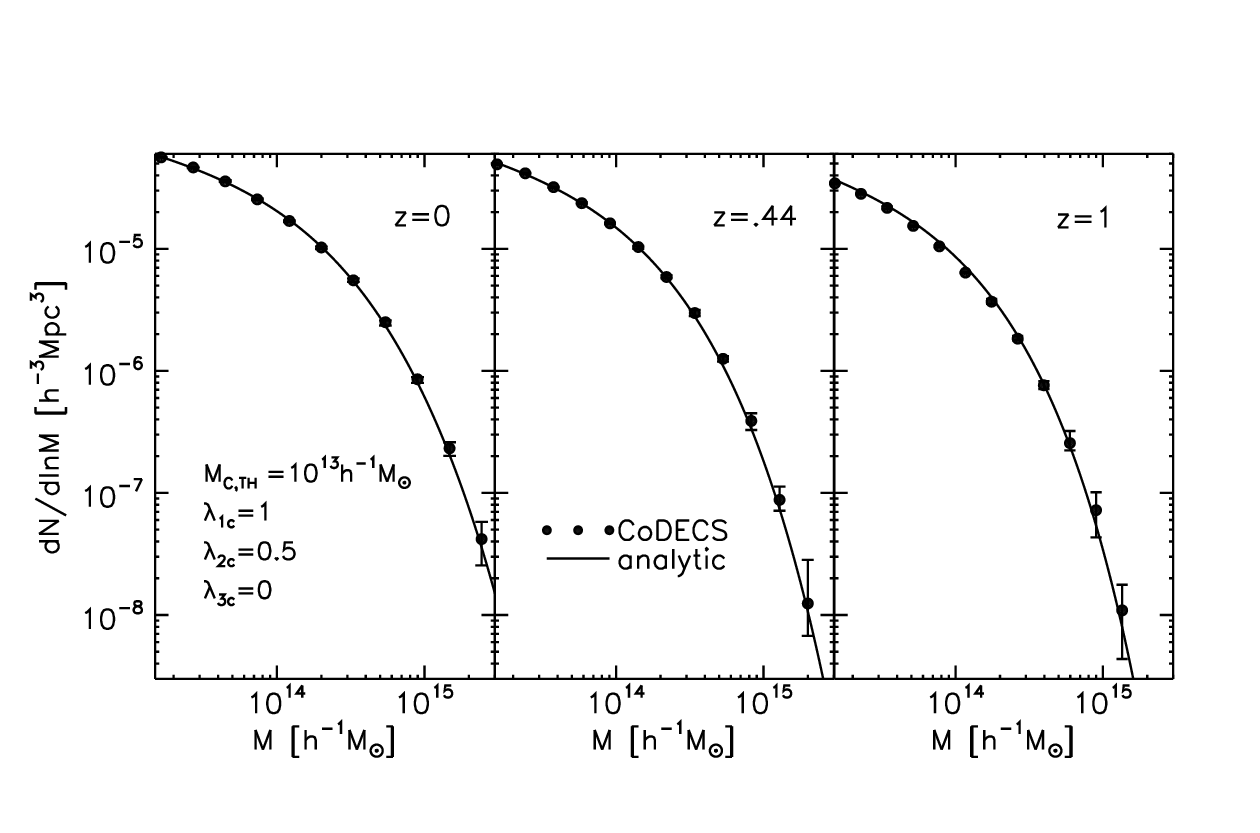}
\caption{Same as Figure \ref{fig:mill} but with the numerical 
results from the CoDECS simulations.}
\label{fig:codecs}
\end{figure}

Due to the relatively small box size of the Millennium simulations 
(see Table \ref{tab:simulation}), however, the numerical results in the high-mass 
section ($M\ge 10^{15}\,h^{-1}M_{\odot}$) suffer from large uncertainties.
Figure \ref{fig:codecs} plots the same as Figure \ref{fig:mill} but for the 
numerical results from the larger CoDECS simulations.  The same values of the two parameters, 
$\lambda_{1c}=1$ and $\lambda_{2c}=0.5$, are consistently 
implemented into our formula while the key cosmological parameters 
are set at the values used for the CoDECS simulations. The analytic supercluster mass functions 
at three redshifts show excellent agreements with the numerical results from the CoDECS 
simulations, too. 

Note that the value of the power spectrum amplitude is different between the two simulations as shown in 
Table \ref{tab:simulation}: $\sigma_{8}=0.9$ for the Millennium while $\sigma_{8}=0.809$ for the CoDECS 
simulations. The excellent agreements between our formula and the numerical results from both 
of the simulations at three different redshifts indicate that the values of the two characteristic 
parameters should be independent of the background cosmology. 

\begin{figure}
\includegraphics[scale=0.85]{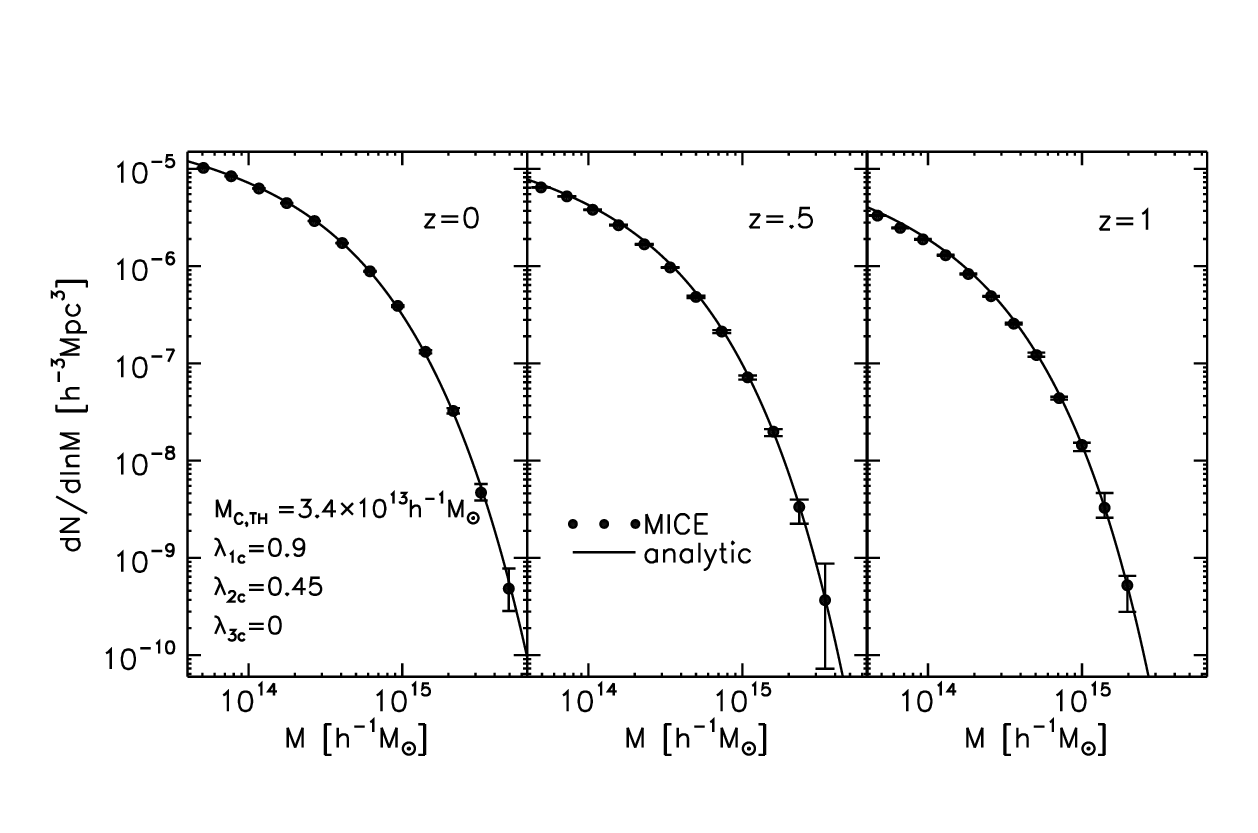}
\caption{Same as Figure \ref{fig:mill} but for the numerical results from the 
MICE simulations. The mass threshold, $M_{\rm c,th}$, for the supercluster 
membership is  $3.4\times 10^{13}\,h^{-1}M_{\odot}$ from the MICE sample.}
\label{fig:mice}
\end{figure}

Figure \ref{fig:mice} plots the same as Figure \ref{fig:mill} but for the numerical results from the 
MICE simulations for which the $\chi^{2}$ statistics yield  lower best-fit values of 
$\lambda_{1c}=0.9$ and $\lambda_{2c}=0.45$. Recall that a higher value of the 
mass threshold, $M_{\rm c, th}=3.4\times 10^{13}\,h^{-1}M_{\odot}$, is used to 
construct the mass-limited sample of the clusters from the MICE simulations. 
This higher value of $M_{\rm c, th}$ results in increasing the mean cluster 
separation, $\bar{l}_{c}$, which should in turn affect the best-fit values of the two parameters. 
Figure \ref{fig:mice} reveals that our formula for the 
supercluster mass function agrees impressively well with the numerical results 
from the MICE simulations, too, even in the high-mass section
($M\ge 3\times 10^{15}\,h^{-1}M_{\odot}$) at all three redshifts. 

Note that the mass-limited cluster sample from the MICE simulation contains less number of the low-mass clusters 
(i.e., the group-size clusters) than those from the other two simulations and thus it has a larger value of the 
mean cluster separation, $\bar{l}_{c}$. In other words, the identified superclusters via the FoF algorithm with the 
fixed linking length of $\bar{l}_{\rm c}/3$ must be less clustered. Henceforth, the values of the characteristic 
parameters of the EZL supercluster mass function, $\{\lambda_{1c},\ \lambda_{2c}\}$, should be related to the 
clustering strength among the member clusters of the superclusters.

\begin{figure}
\includegraphics[scale=0.85]{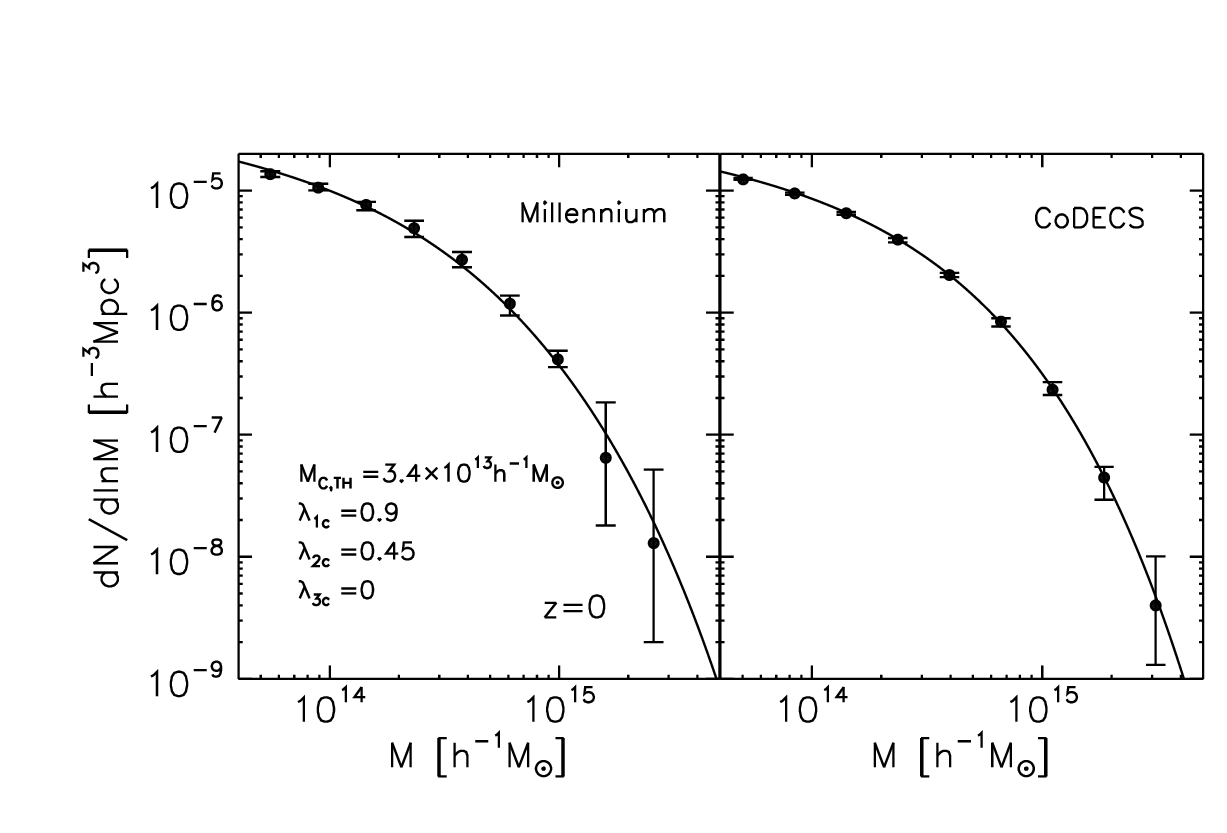}
\caption{Same as Figures \ref{fig:mill} and \ref{fig:codecs} but for the case that 
$M_{\rm c,th}=3.4\times 10^{13}\,h^{-1}M_{\odot}$, the same value used for the 
MICE sample, is adopted for the construction of the mass-limited cluster samples 
from the Millennium simulation (left panel) and from the CoDECS simulations 
(right panel).}
\label{fig:h_mth}
\end{figure}

We would like to examine whether or not the analytic model with the same lower values of 
$\{\lambda_{1c},\ \lambda_{2c}\}$ still agrees well with the numerical results when the same higher 
mass threshold, $M_{\rm c, th}$, is applied to the Millennium and CoDECS cases. 
Figure \ref{fig:h_mth} plots the re-derived numerical results by applying the higher 
mass threshold of $M_{\rm c,th}=3.4\times 10^{13}\,h^{-1}M_{\odot}$ to the Millennium and the 
CoDECS samples and compare them with the analytic formula with  the lower values of the parameters, 
$\lambda_{1c}=0.9$ and $\lambda_{2c}=0.45$. As can be seen, the analytic formula 
still agrees very well with the numerical results from the Millennium and the CoDECS simulations 
even when the higher cluster mass threshold is  adopted. 
This result clearly shows that the values of the characteristic parameters, 
$\{\lambda_{1c},\ \lambda_{2c}\}$, of the EZL supercluster mass function 
depend only on the strength of the clustering in the superclusters but not on the background cosmology.

\section{RELATIVE ABUNDANCE OF THE RICH SUPERCLUSTERS}
\label{sec:rich}

Given that the abundance of the rich superclusters in the universe reflects how fast the structures 
grow and how frequently the clusters merge on the largest scale of the universe, it should be possible 
to use the relative abundance of the rich superclusters as a cosmological probe. To explore this 
possibility, we first define the relative abundance of the rich superclusters as 
\begin{eqnarray}
\label{eqn:rich}
\delta N_{\rm rich}(\ge M_{sc},z)&=&
\frac{N(M\ge M_{sc},z)}{N_{\rm T}(z)}\, , \\
&=&\frac{1}{N_{\rm T}(z)}\int_{M_{\rm sc}}^{\infty}dM\frac{dN(M,z)}{dM},
\end{eqnarray}
where $\delta N_{\rm rich}(\ge M_{sc}, z)$ is the ratio of the cumulative mass function of the 
superclusters with masses larger than $M_{\rm sc}$ to the total number of the 
superclusters, $N_{\rm T}(z)$ at redshift $z$ per unit volume. Note that the relative abundance 
of the rich superclusters is free from the renormalization of the supercluster mass function.

\begin{figure}
\includegraphics[scale=0.85]{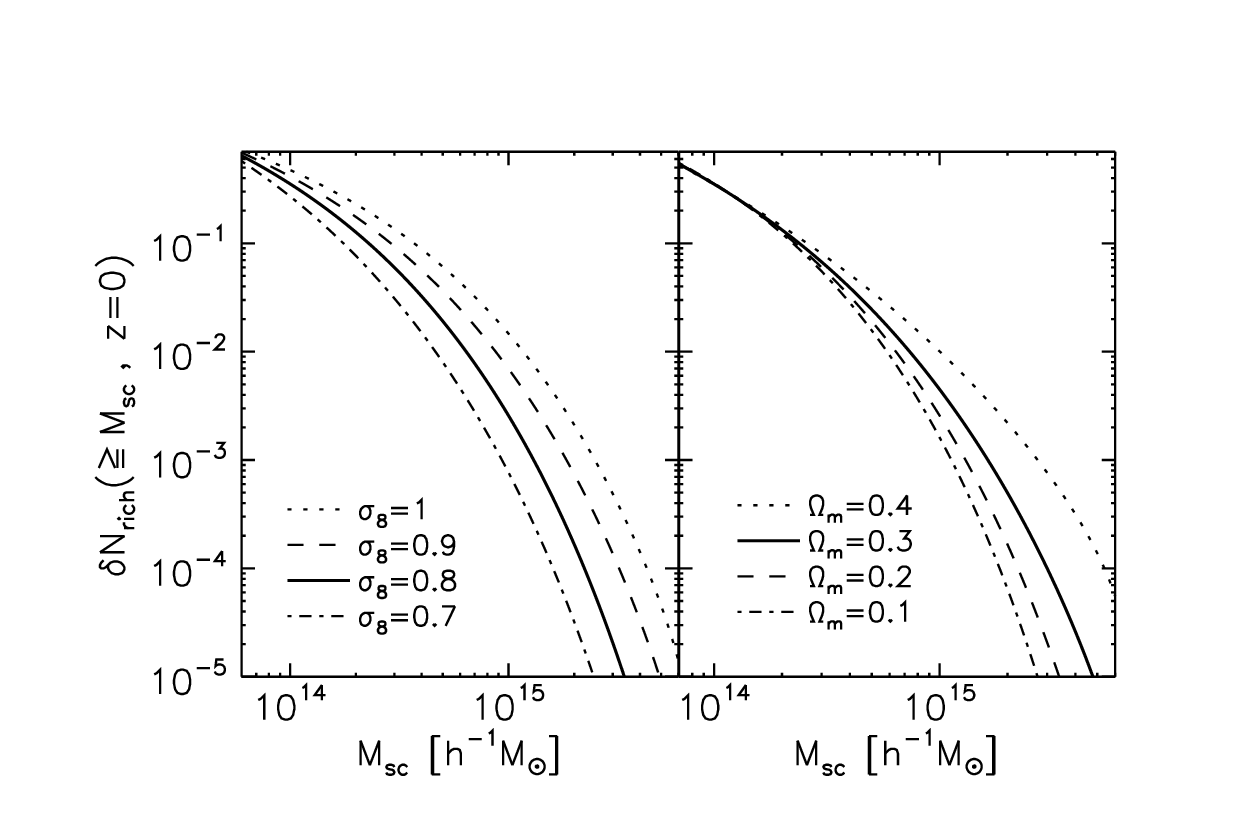}
\caption{Relative abundance of the rich superclusters at $z=0$ for four 
different cases of $\sigma_{8}$ (left panel) and for four different 
cases of $\Omega_{m}$ (right panel).}
\label{fig:rsc_omsig8}
\end{figure}

Using our analytic formula for the supercluster mass function constructed in 
section \ref{sec:model}, we can evaluate $\delta N_{\rm rich}$ as a function of $z$. 
We first investigate the variation of $\delta N_{\rm rich}$ with 
the key cosmological parameters in the standard $\Lambda$CDM model. 
Figure \ref{fig:rsc_omsig8} plots the relative abundance of the rich 
superclusters $\delta N_{\rm rich}$ at $z=0$ as a function of 
$M_{\rm sc} (\ge 10^{15}\,h^{-1}M_{\odot})$ for four different cases of the 
density parameter $\Omega_{m}$ (left panel) and for four different cases of the 
linear power spectrum amplitude $\sigma_{8}$ (right panel). 

For the evaluation of the analytic supercluster mass function, the other 
cosmological parameters are set at the WMAP7 values \citep{wmap7} while 
the characteristic parameters of the supercluster mass function are consistently set at 
$\lambda_{1c}=1$ and $\lambda_{2c}=0.5$.
As can be seen, the variations of $\sigma_{8}$ and $\Omega_{m}$ affect 
significantly the relative abundance of the rich superclusters 
$\delta N_{\rm rich}$ especially in the high mass section. 
As each of $\Omega_{m}$ and $\sigma_{8}$ increases, $\delta N_{\rm rich}$ 
drops less rapidly with $M_{\rm sc}$. This result is consistent with the 
picture that the cosmic web grows faster and the clusters merge more frequently 
in a universe where the initial density fluctuation has higher amplitude and 
dark matter are more dominant.

Assuming that our formula for the supercluster mass function, equation (\ref{eqn:nft}),  
also works in QCDM models (Quintessence+CDM) in which the Quintessence scalar field 
as a dynamical dark energy is responsible for the cosmic acceleration 
\citep[e.g.,][]{caldwell-etal98}, we also perform a feasibility study on how well 
$\delta N_{\rm rich}$ can constrain the dark energy equation of state, $w$, defined 
as $w\equiv P_{Q}/\rho_{Q}$ where $P_{Q}$ and $\rho_{Q}$ represent the pressure 
and density of the Quintessence scalar field, respectively.
For this test, we focus on a toy model in which the dark energy 
equation of state is given as $w(z)=w_{0}+w_{1}z/(1+z)^{2}$ where the two 
parameters, $w_{0}$ and $w_{1}$, have constant values \citep{cp01,lin03}. 
For the evaluation of the supercluster mass function for this toy QCDM model, 
we use the approximate analytic formula for the QCDM linear growth factor 
given in \citet{bas03} \citep[see also][]{per05}. 

\begin{figure}
\includegraphics[scale=0.9]{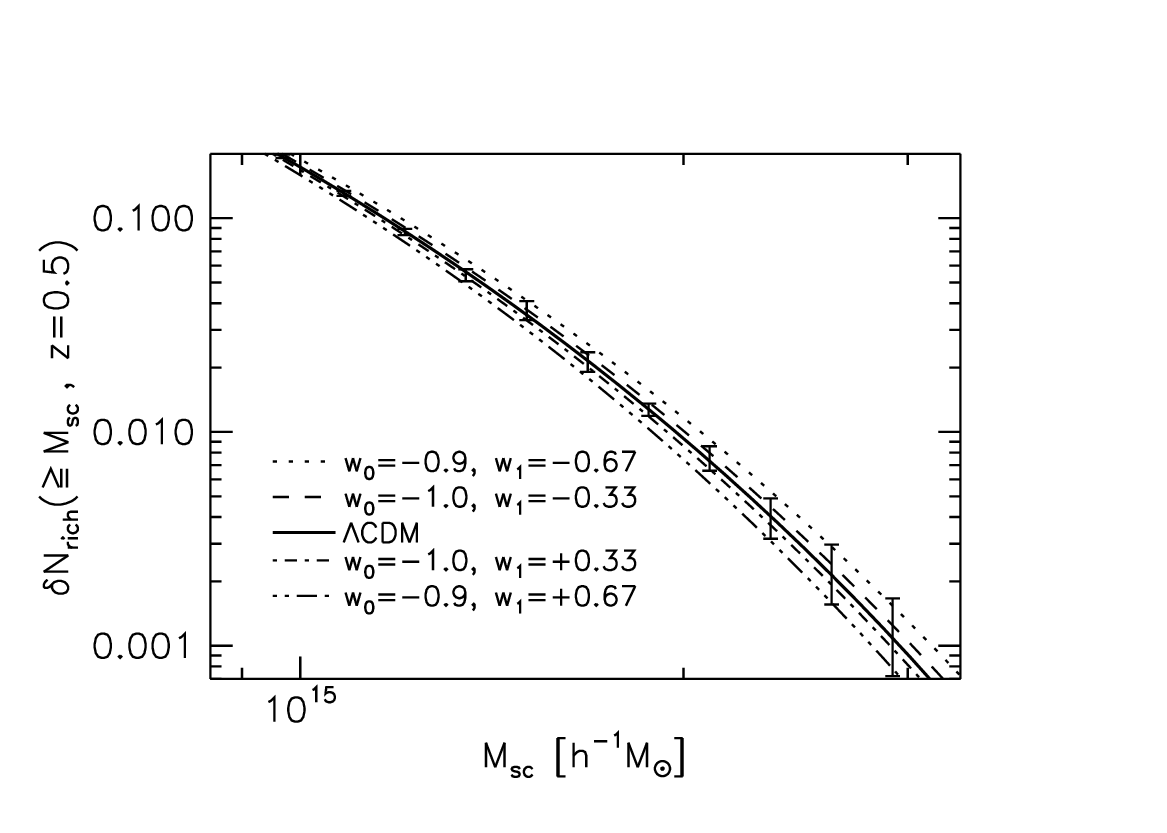}
\caption{Relative abundance of the rich superclusters at $z=0.5$ for  four 
different cases of the dark energy equation of state, 
$w(z)=w_{0}+w_{1}z/(1+z)^{2}$, assuming a toy QCDM model. The fiducial 
$\Lambda$CDM model (with $w_{0}=0,\ w_{1}=0$ and WMAP7 parameters) is also 
shown (solid line) for comparison. The errors are estimated as one standard 
deviation among $8$ Jackknife resamples from the MICE datasets.}
\label{fig:rsc_w}
\end{figure}

Figure \ref{fig:rsc_w} plots the relative abundance of the rich superclusters 
$\delta N_{\rm rich}$ at $z=0.5$ for five different cases of the dark energy 
equation of states. The Jackknife errors from the MICE simulations 
are also overlapped with our models to show explicitly how tight  
the constraints from the relative abundance of the rich superclusters would 
become if the same number of the superclusters were observed in the universe. 
As can be seen, $\delta N_{\rm rich}$ shows an appreciable change with 
$w(z)$ in the high-mass section ($M_{\rm sc}\ge 10^{15}\,h^{-1}M_{\odot}$). 
As the value of $w_{1}$ varies from $-0.67$ to $0.67$, the value of 
$\delta N_{\rm rich}$ decreases by a factor of two on the mass scale of  
$M_{\rm sc}=3\times 10^{15}\,h^{-1}M_{\odot}$, which indicates that the 
abundance of the rich superclusters at a given epoch must be useful to constrain 
the dark energy equation of state.

\begin{figure}
\includegraphics[scale=0.9]{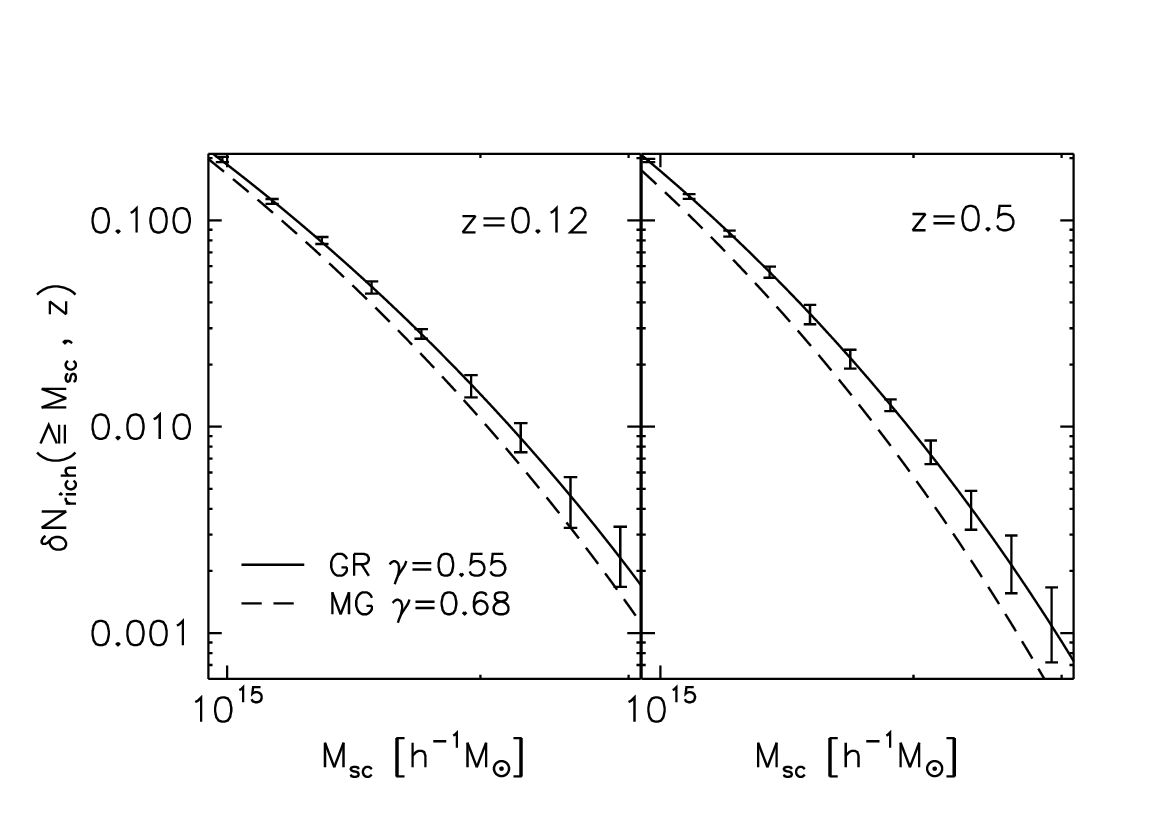}
\caption{Relative abundances of the rich superclusters for the case of the 
fiducial model (GR+$\Lambda$CDM with WMAP7 parameters) (solid line) and for the 
case of a toy model with modified gravity with $D(z)=\Omega^{0.68}_{m}$ 
(dashed line) at $z=0.12$ and $z=0.5$ in the left and right panel, respectively. 
The errors are estimated as one standard deviation scatter among $8$ Jackknife 
resamples from the CoDECS (left) and MICE (right) datasets.}
\label{fig:rsc_g}
\end{figure}

Finally, we also study how $\delta N_{\rm rich}$ changes in a toy modified 
gravity (MG) model in which the linear growth factor scales as a power law of the density 
parameter, $D(z)\propto\Omega_{m}^{\gamma}$  
\citep[e.g., see][]{lin03,lin05}. This toy MG model is 
distinguishable from the standard model (GR+$\Lambda$CDM)  in the 
value of $\gamma$: In the former it is $\gamma=0.68$ while in the latter
it is approximately $\gamma=0.55$ \citep{shapiro-etal10}. Assuming that our formula for 
the supercluster mass function also works in this toy MG model, we 
evaluate $\delta N_{\rm rich}$, which is plotted in Figure \ref{fig:rsc_g} 
(dashed line). The standard (GR+$\Lambda$CDM with WMAP7 parameters) case is 
also plotted with the Jackknife errors for comparison (solid lines) at 
$z=0.12$ and $z=0.5$ in the left and right panels, respectively. The Jackknife 
errors at $z=0.12$ and $z=0.5$ are obtained from the CoDECS and MICE simulations, 
respectively. The difference in $\delta N_{\rm rich}$ between the two models at 
$M_{\rm sc}=3\times 10^{15}\,h^{-1}M_{\odot}$ reaches up to $50\%$ and $66\%$ 
at $z=0.12$ and $0.5$, respectively, which indicates that the relative 
abundance of the rich superclusters at a given epoch must be a useful 
indicator of  modified gravity.

\section{DISCUSSION AND CONCLUSION}\label{sec:con}
 
In the framework of the EZL model constructed in our previous work \citep{LL13}, 
we have provided an efficient formula for the supercluster mass function with two 
characteristic parameters.  The best merit of our formula is that its characteristic parameters 
are robust against variation of the background cosmology and independent of redshifts. 
Extrapolating its validity to non-standard cosmologies, we have suggested that the relative 
abundance of the rich superclusters at a given epoch should be powerful as a cosmological 
probe.  This is the most accurate formula for the supercluster mass function that has ever been 
constructed, achieving a quantitative success in numerical tests. 

Despite the fact that our formula for the supercluster mass function is not a physical one but a merely 
empirical one, the excellent agreements of the formula with the numerical results lead us to expect a 
wide application of the supercluster mass function to various fields. For example,  as mentioned in 
\citet{oguri-etal04}, we expect it to be useful in quantifying how significant the effect of the presence 
of the warm hot intergalactic media on the superclusters \citep{myers-etal04,zappacosta-etal05,SR05}. 
Our formula may also allow us to analytically estimate the late-time integrated Sachs-Wolfe 
(ISW) effect of superclusters \citep[e.g.,][and references therein]{ISW08}.

Before comparing our formula with the supercluster mass function from the real universe, however, 
we will have to undertake a couple of follow-up tasks. The first  task is to examine whether 
or not our formula really works for non-standard cosmologies and investigate how 
its characteristic parameters change when the cosmological constant is replaced by 
the Quintessence scalar field and when there exists a fifth force generated by modified 
gravity. The second task is to account for the projection effect along the directions of 
the line-of-sight on the mass measurement of the superclusters. Since most of 
the superclusters have elongated shapes along the cosmic filaments 
\citep[e.g.,][]{einasto-etal11,tempel-etal12}, 
it would be harder to find the member clusters due to the projection effect if a 
supercluster happens to be elongated along the direction of the line-of-sight. 
We plan to conduct these follow-up works and to report the results elsewhere 
in the future.

\acknowledgments

We thank an anonymous referee who helped us improve significantly the original manuscript.
We acknowledge the use of data from the Millennium, CoDECS and MICE 
simulations that are publicly available at http://www.millennium.com, \\
http://www.marcobaldi.it/web/CoDECS.html, and http://www.ice.cat/mice, 
respectively. 
The Millennium Simulation analyzed in this paper was carried out by the 
Virgo Supercomputing Consortium at the Computing Center of the Max-Planck 
Society in Garching, Germany. The computer code which evaluates the 
mass function of superclusters  will be provided upon request.
This research was supported by Basic Science Research Program through the National Research Foundation 
of Korea(NRF) funded by the Ministry of Education (NO. 2013004372) and partially by the research grant from 
the National Research Foundation of Korea to the Center for Galaxy Evolution Research  (NO. 2010-0027910).

\clearpage
\clearpage
\clearpage
\clearpage
\clearpage
\clearpage
\clearpage
\clearpage

\clearpage
\begin{deluxetable}{cccccc}
\tablewidth{0pt}
\setlength{\tabcolsep}{2.5mm}
\tablecaption{Simulation, linear box size, total number of dark matter 
particles, mass resolution, cosmological parameters and halo-finding 
algorithm.}
\tablehead{Simulation & $L_{\rm box}$ & $N_{\rm p}$ & $M_{\rm p}$ & 
$(\Omega_{m},\ \sigma_{8},\ n)$ & Halo-Finder \\
& [$h^{-1}$Mpc] & & [$10^{8}\,h^{-1}M_{\odot}$] & &} 
\startdata
Millennium & $500$ & $10^{10}$ & $8.6$ & $(0.25,\ 0.9,\ 1)$ & FoF \\
CoDECS  & $1000$ & $2\times1024^3$ & $500.84$ & $(0.271,\ 0.809,\ 0.966)$ & FoF \\ 
MICE & $3072$ & $2048^3$ & $2300.42$ & $(0.25,\ 0.8,\ 0.95)$ & FoF \\  
\enddata
\label{tab:simulation}
\end{deluxetable}
\clearpage
\begin{deluxetable}{cccc}
\tablewidth{0pt}
\setlength{\tabcolsep}{5mm}
\tablecaption{Mass-limited sample, simulation, cluster mass threshold, 
total number of superclusters and mean supercluster mass at $z=0$.}
\tablehead{simulation & $M_{\rm c, th}$ & $N_{\rm tot}$ & 
$\bar{M}_{\rm sc}$ \\
& [$10^{13}h^{-1}M_{\odot}$] & & [$10^{13}\,h^{-1}M_{\odot}$] } 
\startdata
Millennium & $1.0$ & $35422$ & $5.7$ \\
CoDECS & $1.0$ & $412337$ & $6.7$ \\ 
MICE & $3.4$ & $1851534$ & $13.7$ \\ 
Millennium & $3.4$ & $9749$ & $14.0$ \\
CoDECS & $3.4$ & $82016$ & $12.6$ \\
\enddata
\label{tab:sample_z0}
\end{deluxetable}
\clearpage
\begin{deluxetable}{cccc}
\tablewidth{0pt}
\setlength{\tabcolsep}{5mm}
\tablecaption{Same as Table 2 but for $z=0.5$ ($z=0.44$ for CoDECS).}
\tablehead{simulation & $M_{\rm c, th}$ & $N_{\rm tot}$ & 
$\bar{M}_{\rm sc}$ \\
& [$10^{13}h^{-1}M_{\odot}$] & & [$10^{13}\,h^{-1}M_{\odot}$] } 
\startdata
Millennium & $1.0$ & $26333$ & $4.5$ \\
CoDECS & $1.0$ & $333739$ & $5.5$ \\ 
MICE & $3.4$ & $1091086$ & $11.3$ \\ 
\enddata
\label{tab:sample_z0.5}
\end{deluxetable}
\clearpage
\begin{deluxetable}{cccc}
\tablewidth{0pt}
\setlength{\tabcolsep}{5mm}
\tablecaption{Same as Table 2 but for $z=1$.}
\tablehead{simulation & $M_{\rm c, th}$ & $N_{\rm tot}$ & 
$\bar{M}_{\rm sc}$ \\
& [$10^{13}h^{-1}M_{\odot}$] & & [$10^{13}\,h^{-1}M_{\odot}$] } 
\startdata
Millennium & $1.0$ & $17518$ & $3.8$ \\
CoDECS & $1.0$ & $214386$ & $4.4$ \\ 
MICE & $3.4$ &  $494388$ & $9.3$ \\ 
\enddata
\label{tab:sample_z1}
\end{deluxetable}
\end{document}